\documentclass[iop]{emulateapj}
\usepackage{threeparttable}
\usepackage{graphicx}
\usepackage{graphics}
\usepackage{color}           
\usepackage{url}             

\shorttitle{Catastrophic Cooling in Solar TR Loops}
\shortauthors{Orange et al.}

\begin{document}

\title{Direct Observations of Plasma Upflows and Condensation in a Catastrophically Cooling Solar Transition Region Loop}

%
\author{N. B. Orange, D. L. Chesny, H. M. Oluseyi, K. Hesterly, M. Patel, and P. R. Champey}
%
\affil{Department of Physics \& Space Sciences, Florida Institute of Technology, Melbourne, FL  32901}

\begin{abstract}
Minimal observational evidence exists for fast transition region (TR) upflows in the presence of cool loops. Observations of such occurrences challenge notions of standard solar atmospheric heating models, as well as their description of bright TR emission. Using the {\it EUV Imaging Spectrometer} (EIS) onboard {\it Hinode}, we observe fast upflows ($v_\lambda$\,$\le$\,$-$10 km s$^{-1}$) over multiple TR temperatures (5.8\,$\le$\,$\log T$\,$\le$ 6.0) at the footpoint sites of a cool loop ($\log T$\,$\le$\,6.0). Prior to cool loop energizing, asymmetric flows of $+$\,5 km s$^{-1}$ and $-$\,60 km s$^{-1}$ are observed at footpoint sites. These flows speeds and patterns occur simultaneously with both magnetic flux cancellation (at site of upflows only) derived from the {\it Solar Dynamics Observatory}'s (SDOs) { \it Helioseismic Magnetic Imager}'s (HMI) line-of-sight magnetogram images, and a 30\% mass in-flux at coronal heights. The incurred non-equilibrium structure of the cool loop leads to a catastrophic cooling event, with subsequent plasma evaporation indicating the TR as the heating site. From the magnetic flux evolution we conclude that magnetic reconnection between the footpoint and background field are responsible for observed fast TR plasma upflows.
\end{abstract}



\section{Introduction}\label{sec:intro}

A paramount challenge in solar physics still remains --- that is definitively solving the problem of how upper solar atmospheric structures [{\it i.e.}, transition region (TR) and coronal] are heated and maintained \citep{Tripathietal2012ApJ}. Studies of plasma loops, the primary components of each level of the solar atmosphere \citep{Hansonetal1980STIN,Walkeretal1993SPIEa,Walkeretal1993SPIEb,Golubetal1999PhPl,Oluseyietal1999ApJa,Oluseyietal1999ApJb}, have vastly improved and influenced our understanding of solar atmospheric heating, most notably that of the corona \citep{AschwandenNightingale2005ApJ,Warrenetal2008ApJ}.

These basic building blocks of the solar corona, {\it i.e.}, plasma loops, are commonly classified via their peak temperature \citep{Chittaetal2013arXiv}. Hot loops ($\log T$\,$>$\,6.0) have been extensively studied and modeled \citep[{\it e.g.},][]{Mackayetal2010SSRv,AschwandenSchrijver2002ApJS,Spadaroetal2006ApJ}, while to a much lesser extent diffuse cool (TR) loops \citep[$\log T$\,$\le$\,6.0; {\it e.g.},][]{Chittaetal2013arXiv,Tripathietal2012ApJ,Chesnyetal2012,Mulleretal2003A&A,Mulleretal2004A&A,Oluseyietal1999ApJa,Oluseyietal1999ApJb}. Recent observational and theoretical advances on the heating of plasma confined within hot loop structures have revealed that both steady-state \citep[{\it e.g.},][]{Winebargeretal2011ApJ,Warrenetal2010ApJ} and impulsive heating \citep[{\it e.g.},][]{ViallKlimchuk2012ApJ,Tripathietal2010ApJ} processes are consistent with their observed temperature and intensity structures. Impulsive heating has been found to explain the properties of cool loops \citep[{\it e.g.},][]{Spadaroetal2003ApJ}, that are not in equilibrium and constantly evolving \citep[{\it e.g.},][]{Ugarteetal2009ApJ}. However, what remains unknown is the role plasma condensation plays in such processes \citep{Chittaetal2013arXiv}.

Impulsive heating events, bundles of nanoflare heated loop strands, occur at coronal heights and result in chromospheric evaporation  \citep[{\it e.g.,},][]{Klimchuketal2008ApJ,Klimchuk2009ASPC}. Bright TR emission is widely considered a response to cooling coronal plasma which was impulsively heated. The pervasively observed TR redshifts \citep{Hansteenetal1996ASPC} and observations that cool loops are characterized by plasma downflows, at footpoints and along the loop structures
\citep[{\it e.g.},][]{DelZanna2008A&A,Tripathietal2009ApJ}, have provided significant support to such models. However, observational evidence is emerging (to our knowledge only those reported by \citeauthor{Tripathietal2012ApJ} \citeyear{Tripathietal2012ApJ}) that reveals the existence of fast TR upflows in the presence of cool loops. The significance of such results is the introduction of challenges to standard solar atmospheric heating models. Particularly these are their descriptions on the origin of bright TR emission and heights at which coronal heating occurs. Furthermore, emerging observational evidence for fast TR upflows, mainly associated with explosive events \citep{Beckers1968SoPh,Beckers1968SoPhB} and spicules \citep{DePontieuetal2007PASJ,Langangenetal2008ApJ}, are providing increasing support that atmospheric heating is not confined to the corona.

Though observational evidence exists indicating heating occurs in cooler regions of the solar atmosphere; the heights, timescales, and mechanisms responsible remain unclear \citep{Tripathietal2012ApJ}. Moreover, a current topic of hot debate is whether fast TR upflows provide significant mass-influx to coronal heights, as well as the role they play in the generation of observed coronal phenomena \citep{DePontieuetal2009ApJ,Langangenetal2008ApJ,Klimchuk2012JGRA}. It is also emphasized that minimal investigations have been carried out on how plasma condensation relates to non-thermal equilibrium states of cool loops \citep{Mulleretal2003A&A,Mulleretal2004A&A}, while the role of the underlying magnetic field plays in such scenarios remains unquantified \citep{Chittaetal2013arXiv}.

In relation to the above discussion, we have on hand a unique data-set derived from observations taken with the {\it EUV Image Spectrometer} \citep[EIS;][]{Culhaneetal2007SoPh} onboard {\it Hinode} that provides direct observational evidence of high-speed upflows at multiple TR temperatures occurring at the footpoint sites of a catastrophically cooling loop. We compliment these data with line-of-sight (LOS) magnetogram observations from the {\it Heliosemic and Magnetic Imager} \citep[HMI;][]{Schouetal2012SoPh} onboard {\it Solar Dynamics Observatory} (SDO) to investigate the effects of magnetic flux evolution on both plasma upflows and the runaway cooling event.

In that respect the rest of the paper is as follows: observational data its processing and analysis are presented in the next section, $\S$~\ref{sec:results} presents the measurement results, the discussion of these results and our conclusions are provided in $\S$'s~\ref{sec:discussion} and \ref{sec:conclusions}, respectively.

\section{Observations, Processing, and Analysis}\label{sec:processing}
\begin{figure}[h]
\centering
\includegraphics[scale=0.3]{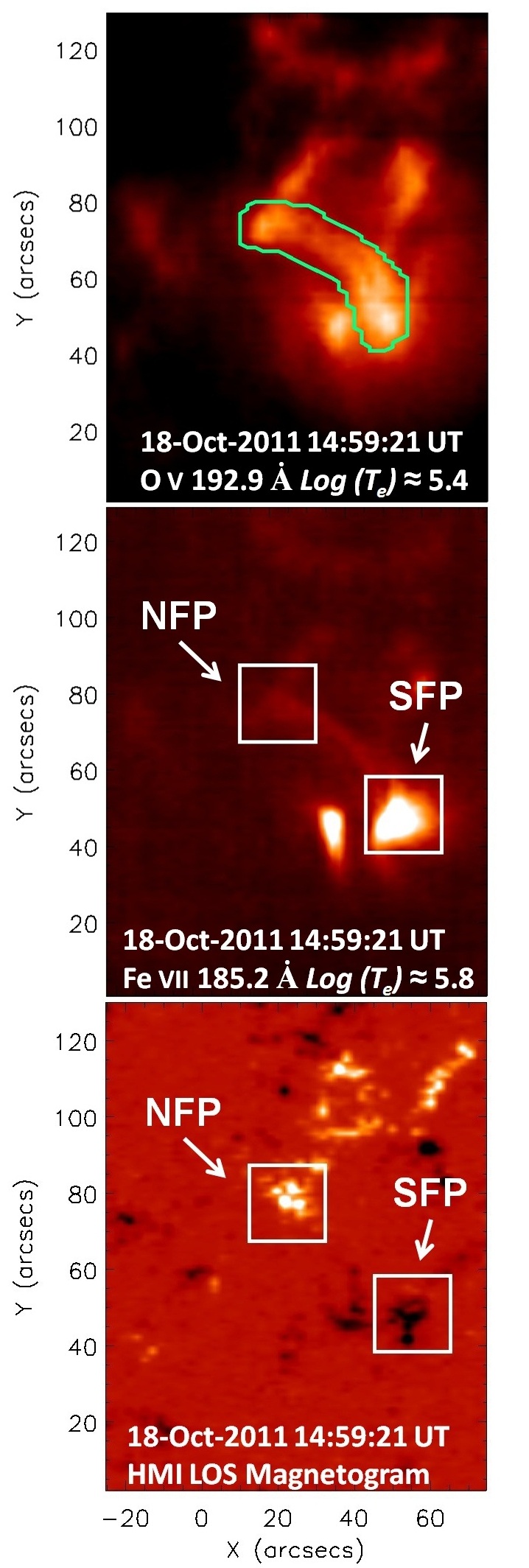}
   \caption{Top and middle: EIS 2$^{\prime \prime}$ intensity images (erg cm$^{-2}$ s$^{-1}$ sr$^{-1}$), observed 18 October 2011 at 14:59:21 UT, of O\,\textsc{v} (192.90\,{\AA}, $\log T$\,$\approx$\,5.4) and Fe\,\textsc{viii} (185.20\,{\AA}, $\log T$\,$\approx$\,5.8), respectively, with green contours denoting the loop and boxed regions denoting it's footpoints, respectively. Bottom: HMI LOS magnetogram (G) displayed over the FOV of the EIS observation observed 18 October 2011 at 14:59:21 UT with footpoint regions again identified.
   }\label{fig:LOOP_FPS_Marked}
\end{figure}

Observational data was obtained from EIS during the following time frame: 14:02:36 UT to 15:19:55 UT at $\approx$\,20 min time intervals on 18 October 2011. The data consists of raster scans with a 2$\arcsec$ slit width using 1$\arcsec$ steps resulting in a final field-of-view (FOV) of 100$\arcsec$\,$\times$\,130$\arcsec$ (Figure~\ref{fig:LOOP_FPS_Marked}). Nine emission lines providing temperature coverage from the chromosphere ($\log T$\,$\approx$\,4.9) to the corona ($\log T$\,$\approx$\,6.4) were used. In Table~\ref{tbl:eis_elines} we provide the emitting ions, their respective wavelengths, and peak formation temperatures.

\begin{table}[t]
\centering
\caption{List of EIS observed emission lines with format of: Ion, wavelength ({\AA}), and logarithmic electron temperature.} \label{tbl:eis_elines}
\begin{tabular}{ccc}
\hline
Ion & Wavelength (\AA) & $\log T$ \\
\hline
He \textsc{ii}   & 256.32 & 4.9 \\
O \textsc{v}     & 192.90 & 5.4 \\
Fe \textsc{viii} & 185.21 & 5.8 \\
Fe \textsc{ix}   & 197.86 & 5.9 \\
Fe \textsc{x}    & 184.54 & 6.0 \\
Fe \textsc{xii}  & 186.88 & 6.2 \\
Fe \textsc{xii}  & 195.12 & 6.2 \\
Fe \textsc{xiv}  & 274.20 & 6.3 \\
Fe \textsc{xv}   & 284.16 & 6.4 \\
\hline
\end{tabular}
\end{table}

Image pre-processing of EIS level-0 data was performed with standard {\sf Solar SoftWare} (SSW) to obtain flux calibrated data. Additional corrections were made for the spatial offset occurring between its short (171--212\,{\AA}) and long (250--290\,{\AA}) wavelength bands \citep{YoungGallagher2008SoPh}, instrument and orbital jitter variations \citep{Shimizuetal2007}, CCD spectrum drift \citep{Mariskaetal2007PASJ}, and tilt of the emission on the detector.
The resultant EIS level-1 image data then possesses an absolute wavelength calibration of $\pm$\,4.4 km s$^{-1}$ \citep{Kamioetal2010SoPh}.

\begin{figure*}[!t]
\centering
\includegraphics[scale=0.35]{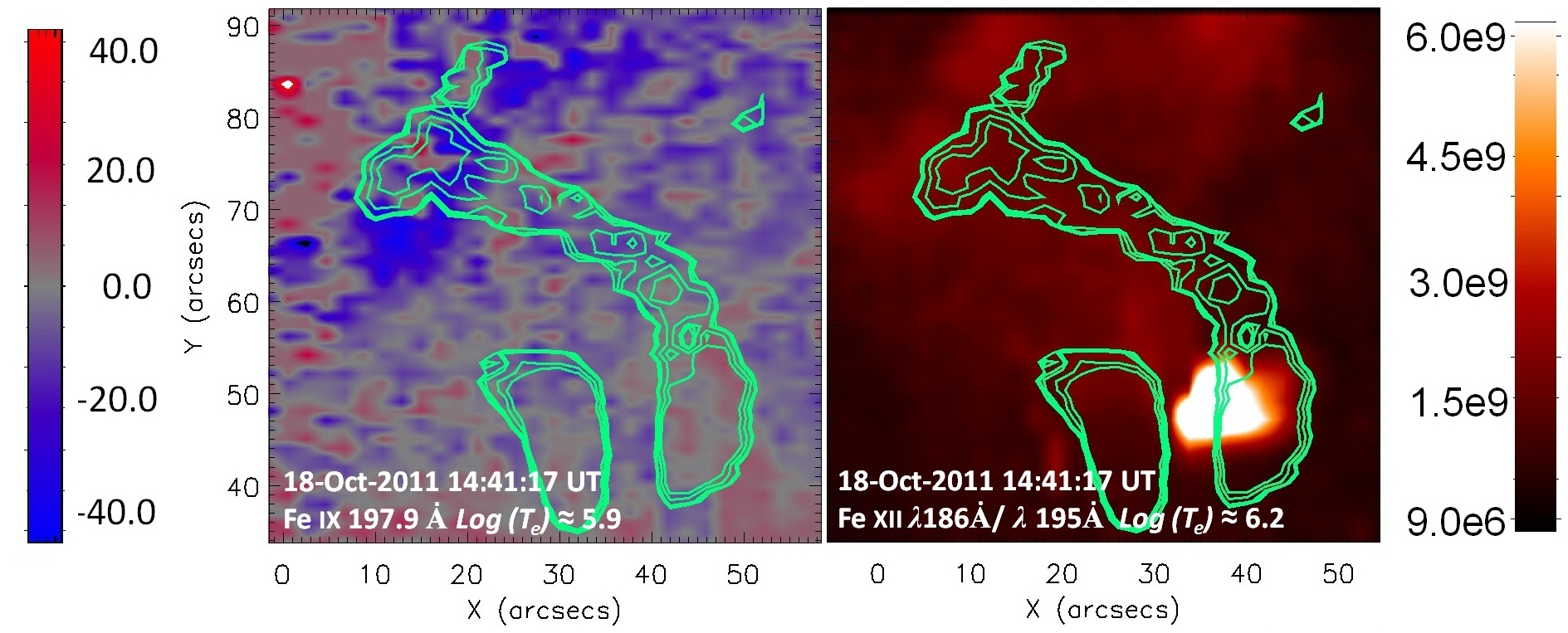}
   \caption{{\it Top and bottom:} LOS velocity in (km s$^{-1}$) and electron density (in cm$^{-3}$) images, both observed at 14:41:17 UT on 18 October 2011, derived from the intensity image of the Fe \textsc{ix} (197.86 {\AA}, $\log T$ $\approx$ 5.9), and Gaussian intensity ratio of Fe\,\textsc{xii}\,186\,{\AA}/195\,{\AA}, respectively. On both panels green contours are intensity levels of 80\%\,--\,90\% of the peak intensity derived from Fe \textsc{ix} emission line observed at 14:59:21 UT denoting the complete filled cool loop.
   }\label{fig:C1_fe9Vel_Fe9Intover}
\end{figure*}

\begin{figure*}[!ht]
\centering
\includegraphics[scale=0.59]{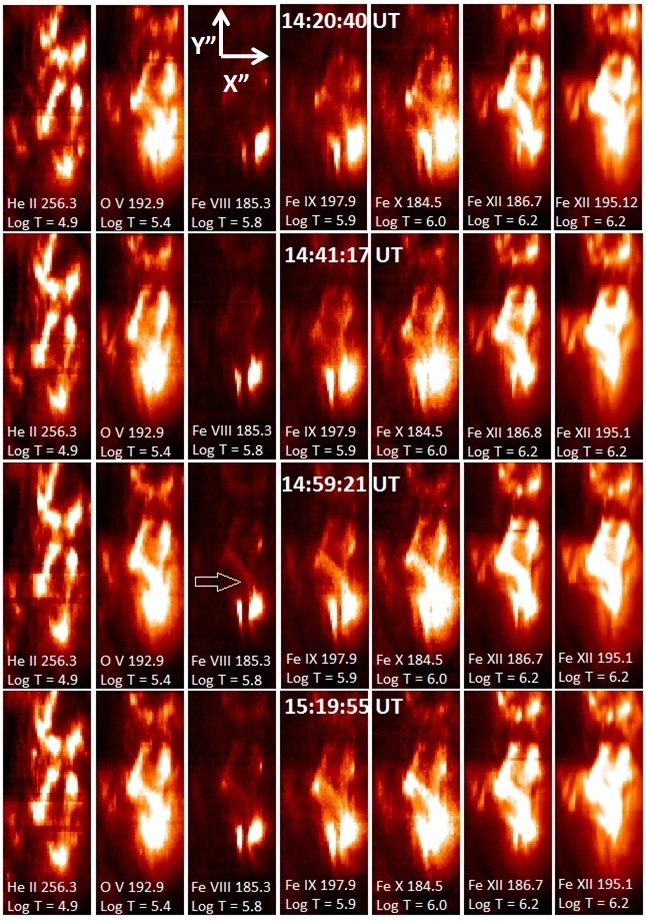}
   \caption{EIS intensity images (erg cm$^{-2}$ s$^{-1}$ sr$^{-1}$), covering electron temperatures in the range of $\log T$\,$\approx$\,4.9\,--\,6.2 (left to right, respectively), observed on 18 October 2011 from 14:20:40 UT to 15:19:55 UT (top to bottom, respectively) showing both the cool ($\log T$\,$\le$\,6.0) and hot loop ($\log T$\,$>$\,6.0) flux evolution. Note, an arrow identifies the filled cool loop corresponding to the time of peak emission in the core region.
   }\label{fig:FFOV_1420_1459_snapshot}
\end{figure*}

The standard SSW routine {\sf eis\_auto\_fit.pro} \citep{Young2010} was used to derive integrated spectral line intensities and their respective Doppler shifts, as well as build both intensity and LOS velocity images of the loop (Figures~\ref{fig:LOOP_FPS_Marked} and \ref{fig:C1_fe9Vel_Fe9Intover}). This routine fits a single Gaussian to each pixel forming the raster scan with the well-known {\sf mpfit.pro} algorithm and propagates uncertainties from 1$\sigma$ fit uncertainties. However we note, during this process multiple Gaussian fits were applied to both the He\,{\sc ii}\,256.32\,{\AA} and O\,{\sc v}\,192.90\,{\AA} spectrums due to their blending with coronal emission lines. Particulary, He\,{\sc ii}\,256.32\,{\AA} is blended with that of the Si\,{\sc x}\,256.37\,{\AA} and Fe\,{\sc x}\,256.41\,{\AA} lines, while O\,{\sc v}\,192.90\,{\AA} is blended with that of the Fe\,{\sc xi}\,192.83\,{\AA} line \citep{Youngetal2007PASJ857,Brownetal2008ApJS}. Visual inspection of He\,{\sc ii}\,256.32\,{\AA} fits were consistent with the report of \citet{DelZanna2012arXiv} that in on-disk quiet Sun regions it contributes over 80\% of the observed intensity. Further support of this notion is found in a direct comparison of distinct bright network regions of the Fe\,{\sc x} and He\,{\sc ii} intensity images in Figure~\ref{fig:FFOV_1420_1459_snapshot}. O\,{\sc v} spectrum clearly indicated, and supported notions, that in most regions the Fe\,{\sc xi}\,192.83\,{\AA} dominates \citep{Youngetal2007PASJ727}, which is also observed in the similarity of O\,{\sc v} intensity images of Figure~\ref{fig:FFOV_1420_1459_snapshot} to those of the corona. However, in the cool loop regions ({\it i.e.}, such as those denoted by contours and boxes on Figure~\ref{fig:LOOP_FPS_Marked}) enhanced O\,{\sc v}\,192.90\,{\AA} emission was observed, and as such allowed it's spectrum to be resolved and fitted. Moreover, we point out \citet{Youngetal2007PASJ727} observed similar physical characteristics in the spectrum around 192.83\,{\AA} for a study of an active region TR brightening.

Full disk LOS magnetograms from HMI are utilized to investigate the magnetic structure of the loops. HMI's magnetograms were obtained with a spatial resolution of $\approx$\,0$\farcs$5 at a cadence of $\approx$\,45 s and corresponded to the approximately one hour of EIS observations. Magnetogram data was pre-processed using standard SSW techniques incorporating per-pixel noise subtraction \citep{Brownetal2011} and then averaged over $\approx$\,2.5 min intervals to increase the signal-to-noise ratio with additional pointing corrections using the techniques of \citet{Orangeetal2012b}. Magnetograms were co-aligned to EIS scans, with observational time differences of $\lesssim$\,three minutes, using the SSW routine {\sf drot\_map.pro}. A resultant alignment error of $\lesssim$\,2$\arcsec$ was measured  by cross-correlating visually bright coronal structures to strong magnetogram regions. We also note, the image's vicinity to solar disk center ($\lesssim$\,$\pm$\,150$\arcsec$ in both the solar $x$ and $y$ directions) result in negligible projection effects.

The loop plus minimal background emission was defined by using a semi-supervised tracing algorithm applied at each observational time step to O\,\textsc{v}\,192.9\,{\AA} ($\log T$\,$\approx$\,5.4) intensity images (Figure~\ref{fig:LOOP_FPS_Marked}). The resultant region was then used to isolate the loop in all other EUV images. We assigned the loop a set coordinates $s$, corresponding to pixels along its spine, and segmented them into three distinct regions of core, north, and south footpoints (NFP and SFP, respectively; Figure~\ref{fig:LOOP_FPS_Marked}). Segmentation was performed using loop footpoint regions identified in Fe\,\textsc{viii}\,185.2\,{\AA} ($\log T$\,$\approx$\,5.8) intensity images at each time step. It is noted, the footpoint regions were visually verified in corresponding He\,\textsc{ii}\,256.3\,{\AA} ($\log T$\,$\approx$\,4.9).

Prior to measuring radiative flux ($F_{\lambda}$; arbitrary units) as a function of loop length $s$, we apply a rigorous background subtraction method. In this method, solar background/foreground emission is removed from the loop structure by applying custom written software to image thumbnails with the loop at their center and background/foreground emission surrounding it. The technique obtains a background estimate by: first, applying a low-pass filter to remove high-frequency noise such as bad pixels; then, uses a histogram of image flux to isolate the lowest 10\%; and finally, applies a weighted average to the lowest 98\% of the isolated flux. It is noted, care is taken to reduce background overestimation (i.e., generation of negative pixel values) since previous studies have shown that coronal intensities of loop structures are $\approx$\,10\% -- 20\% higher than background/foreground emission
\citep[{\it e.g.},][]{ViallKlimchuk2012ApJ,DelZannaMason2003A&A}. The result is a background subtracted image where pixel values are reduced by $\lesssim$\,10\%. We then measure $F_{\lambda}(s)$ by averaging over cross-sections of loop width at each point along its spine. Light curves are generated of total radiative flux for each respective region ({\it i.e.}, loop core, NFP, and SFP) by integrating their 3$\sigma$ brightest flux.

Resultant spectral intensities of the Fe\,\textsc{xii} emission lines (Table~\ref{tbl:eis_elines}) were used to generate electron density images of the loop (Figure~\ref{fig:C1_fe9Vel_Fe9Intover}) via the techniques discussed by \citet{Young2011}. It is noted, though each Fe\,\textsc{xii} emission line is blended, previous studies have suggested for densities $<$\,10$^{10}$ cm$^{-3}$, consistent with results herein, blending effects to the Fe \textsc{xii}\,195\,{\AA} line are not important \citep{Dere2008A&A}. Moreover, \citet{Dereetal2007PASJ} have suggested a root-mean-square error of $\approx$\,1.6 for determination of densities in the quiet Sun when this line ratio is utilized in the aforementioned conditions. Then, LOS velocities ($v_{\lambda}$; km s$^{-1}$) and electron density ($N_{{\rm e}}$; cm$^{-3}$) are measured as a function of loop length by the techniques described above. Light curves of these parameters as a function of loop region are obtained by smoothing over the core and footpoint regions as function of time.
\begin{figure*}[!t]
\centering
\includegraphics[scale=0.38]{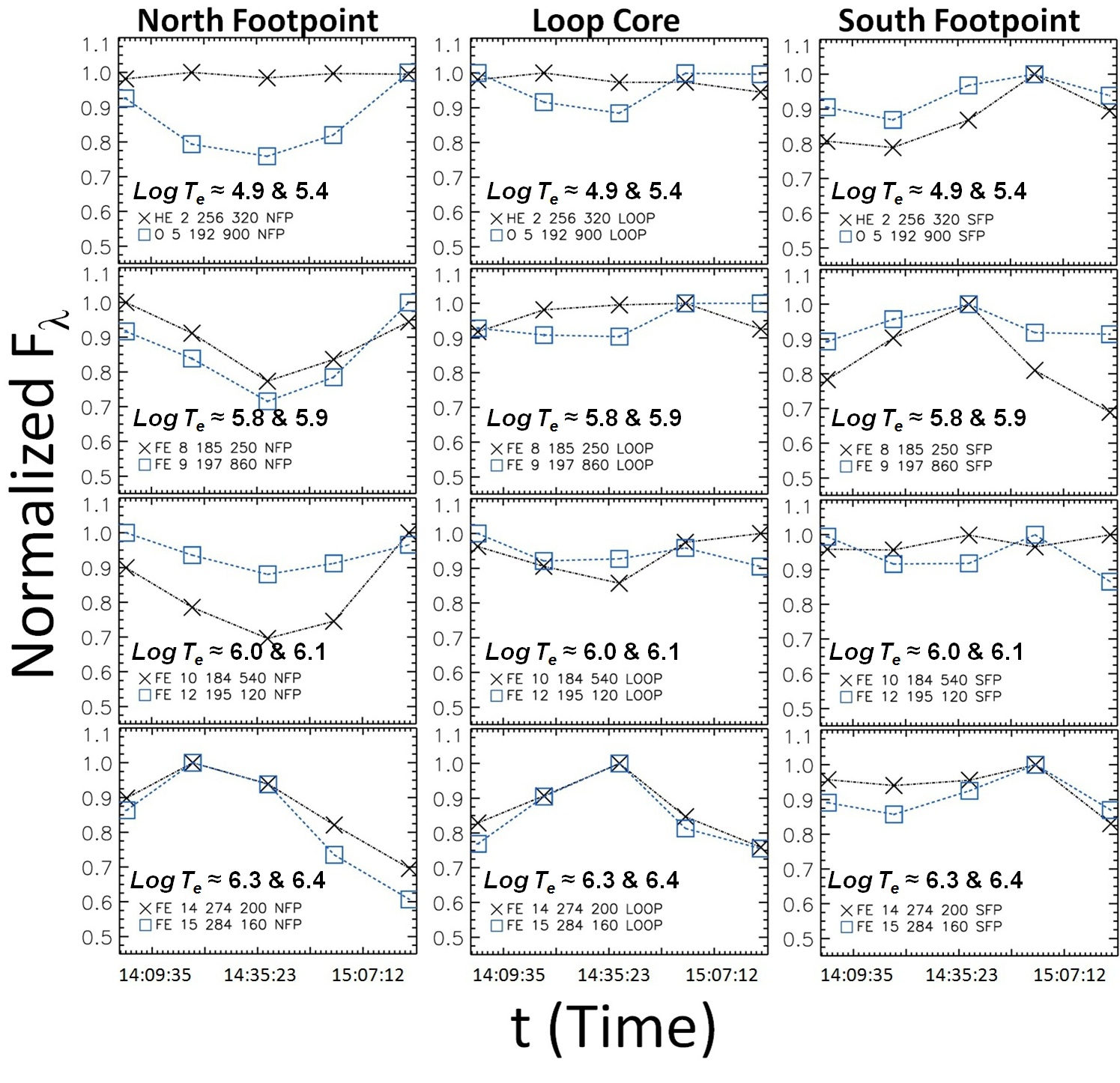}
   \caption{Flux $F_\lambda$ (arbitrary units) vs time (14:02 UT\,--\,15:19 UT) of the loop's NFP, core, and SFP regions (left to right, respectively) displayed from top to bottom as a function of increasing temperature for the emission lines of Table~\ref{tbl:eis_elines}, respectively.
   }\label{fig:NFP_CORE_SFP_LCS}
\end{figure*}

The loop's footpoint regions are used to aggregate magnetic field data at each time step (Figure~\ref{fig:LOOP_FPS_Marked}). These magnetogram data cubes are used to measure the magnetic flux density, {\it i.e.}, the number of positive and negative polarity elements above and below a threshold value of 20 G, respectively. This threshold value is consistent with the average field-of-view strength of our magnetic field imagery. It is noted, analyzing magnetic field imagery in this manner provides information on flux likely contributing to reconnection events given the notion that stronger flux ({\it i.e.}, $>$ threshold) reconnects while weaker flux ({\it i.e.}, $<$ threshold) is ``scattered" (\citeauthor{Sakaietal1997SPD} \citeyear{Sakaietal1997SPD}; Chesny 2013, private communication).

To examine the temperature structure as a function of loop region, we use the aforementioned measurements of total radiative flux to execute an emission measure EM loci analysis via the techniques discussed by \citet{Orangeetal2012a}. We employ the physical assumptions used by \citet{Orangeetal2012a}, with exception of $N_{{\rm e}}$ which is derived from our measurements discussed above. Note, loci curves represent the EM as it originates from isothermal plasma at a given temperature thereby revealing isothermal plasmas where all curves meet \citep{Kamioetal2011A&A}.

\section{Results}\label{sec:results}

Shown in the top row of Figure~\ref{fig:FFOV_1420_1459_snapshot} and corresponding to the 14:20:40 UT, temperatures $\le$\,1 MK are characterized by visually bright footpoints and diffuse partially to not filled loops. The cool loop can then be seen beginning to fill from the SFP to the NFP at $\log T$\,$\approx$\,5.8 and 14:41:17 UT ({\it i.e.}, 2$^{nd}$ row from top of Figure~\ref{fig:FFOV_1420_1459_snapshot}). The loop is completely filled ($\log T$\,$\le$\,6.0), with a typical length $\approx$\,40 Mm, by 14:59:21 UT ({\it i.e.}, 3$^{rd}$ row from top Figure~\ref{fig:FFOV_1420_1459_snapshot} and identified by arrow on Fe\,\textsc{viii}\,185.2,{\AA} image). In the last observation time, 15:19:55 UT, the loop is returning to equilibrium and has cooled sufficiently given the observed decreases in visually bright plasma.

In Figure~\ref{fig:NFP_CORE_SFP_LCS} we have provided the light curves of the NFP, core, and SFP. In the NFP, the flux decreases until the cool loop begins to fill (14:41 UT) and increases thereafter, with exception of $\log T$\,$\ge$\,6.2 which continually decrease. The core region of the loop experiences similar trends in flux evolution, as function of temperature, as that of the NFP, with exception  $\log T$\,$\ge$\,6.2 which exhibit a triangular shape. SFP light curves for temperatures over 5.8\,$\le$\,$\log T$\,$\le$\,6.0 peak in irradiance at 14:41 UT, while both cooler and hotter regimes peak $\approx$\,20 min later (14:59 UT).

In relation to the cool loop, $\log T$\,$\le$\,6.0, irradiance peaks are consistent with the structural evolution, observed in Figure~\ref{fig:FFOV_1420_1459_snapshot} ({\it i.e.}, consistent with previous results that loop fills from the SFP to the NFP). In terms of the temporal evolution these flux peaks occur as follows: first SFP at 14:41 UT, then core at 14:59 UT, and finally NFP at 15:19 UT. Inspection of Figure~\ref{fig:NFP_CORE_SFP_LCS} indicates that the cool loop is not a result of cooling coronal material, based on comparisons between peak TR and coronal fluxes at the SFP site where said TR flux peaks precede those of hotter temperatures. Furthermore, the SFP peak in TR EUV flux transverses the loop back to the NFP while hotter regions peak in the NFP and progress towards the SFP site.

\begin{figure}[!t]
\centering
\includegraphics[scale=0.3]{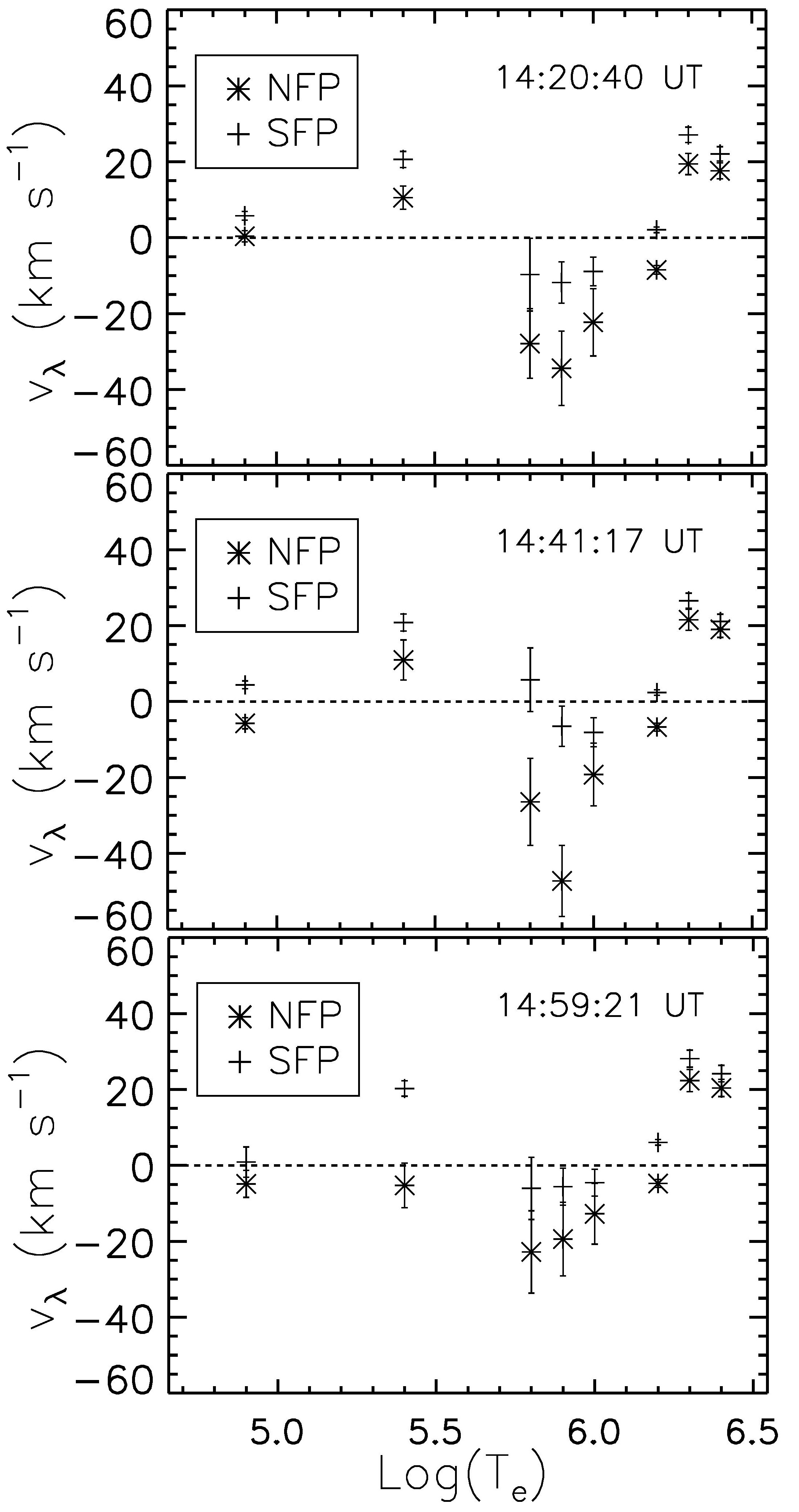}
   \caption{LOS velocity (km s$^{-1}$) versus electron temperature ($\log T$) for the NFP (asterisks) and SFP (pluses) regions derived from observational data  on 18 October 2011 at the observational times of 14:20:40 UT\,--\,14:59:21 UT (top to bottom, respectively). Note, $v_{\lambda}$\,$<$\,0 and $>$ indicate upflows and downflows, respectively, while $v_{\lambda}$\,$=$\,0 is denoted by dashed line.
   }\label{fig:FPS_1420_LOSV_VS_TE}
\end{figure}

At the NFP prior to the appearance of the cool loop (14:20 UT; $\approx$\,40min), temperatures over 5.8\,$\le$\,$\log T$\,$\le$\,6.2 were characterized by plasma upflow speeds of $\approx$\,8.0 -- 40 km $s^{-1}$ (Figure~\ref{fig:FPS_1420_LOSV_VS_TE}) with plasma falling at hotter and cooler temperatures. In the NFP at 14:41 UT and $\log T$\,$\approx$\,5.9 the plasma upflow speed peaked at $\approx$\,-60 km s$^{-1}$, with upflowing plasma still occurring over a temperature range of 5.8\,$\le$\,$\log T$\,$\le$\,6.2. However, at the SFP site and this same temperature range plasma was falling at a typical rate of $\lesssim$\,5 km s$^{-1}$  (Figure~\ref{fig:FPS_1420_LOSV_VS_TE}). Upon complete filling of the cool loop ({\it i.e.}, 14:59 UT) the plasma flow directions versus temperature had returned to their typical form, that is upflowing plasma in the upper TR and lower corona (5.8\,$\le$\,$\log T$\,$\le$\,6.2) with falling plasma at cooler and hotter temperatures. We point out, maximized NFP plasma upflows, particularly over 5.8\,$\le$\,$\log T$\,$\le$\,6.0 temperatures, corresponded with the transition from upflows to downflows, at similar temperatures, in the SFP region.

\begin{figure}[!t]
\centering
\includegraphics[scale=0.31]{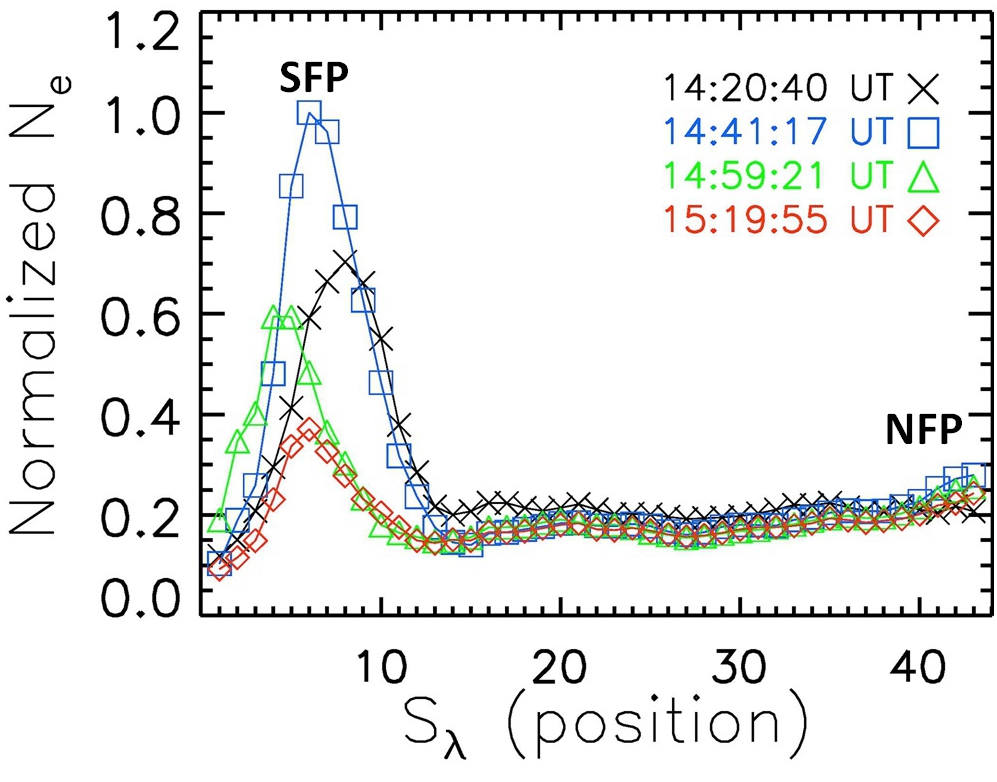}
   \caption{Normalized electron density ($\log N_{{\rm e}}$, at $\log T$\,$\approx$\,6.2) as a function of loop position ($s_\lambda$). The NFP and SFP regions are denoted on the plot, while the observation times of 14:20:40 UT, 14:41:17 UT, 14:59:21 UT, and 15:19:55 UT are represented by x's, squares, triangles, and diamonds, respectively.
   }\label{fig:1420_1459_densevol}
\end{figure}

The coronal ($\log T$\,$\approx$\,6.2) electron density evolution in the NFP was roughly constant over the time frame studied here (Figure~\ref{fig:1420_1459_densevol}). However, in the SFP region significant density fluctuations occurred and are described as follows (Figure~\ref{fig:1420_1459_densevol}). During the first $\approx$\,20 mins (14:20 UT\,--\,14:41 UT) a mass in-flux of $\approx$\,30\% is found. Next, mass-loss of $\approx$\,40\% is witnessed over the next $\approx$\,20 mins (14:41 UT\,--\,14:59 UT). Finally, during the time in which the loop cooled completely (14:59 UT\,--\,15:19 UT) it continued to lose mass with a total loss of $\approx$\,30\%.

\begin{figure*}[!t]
\centering
\includegraphics[scale=0.4]{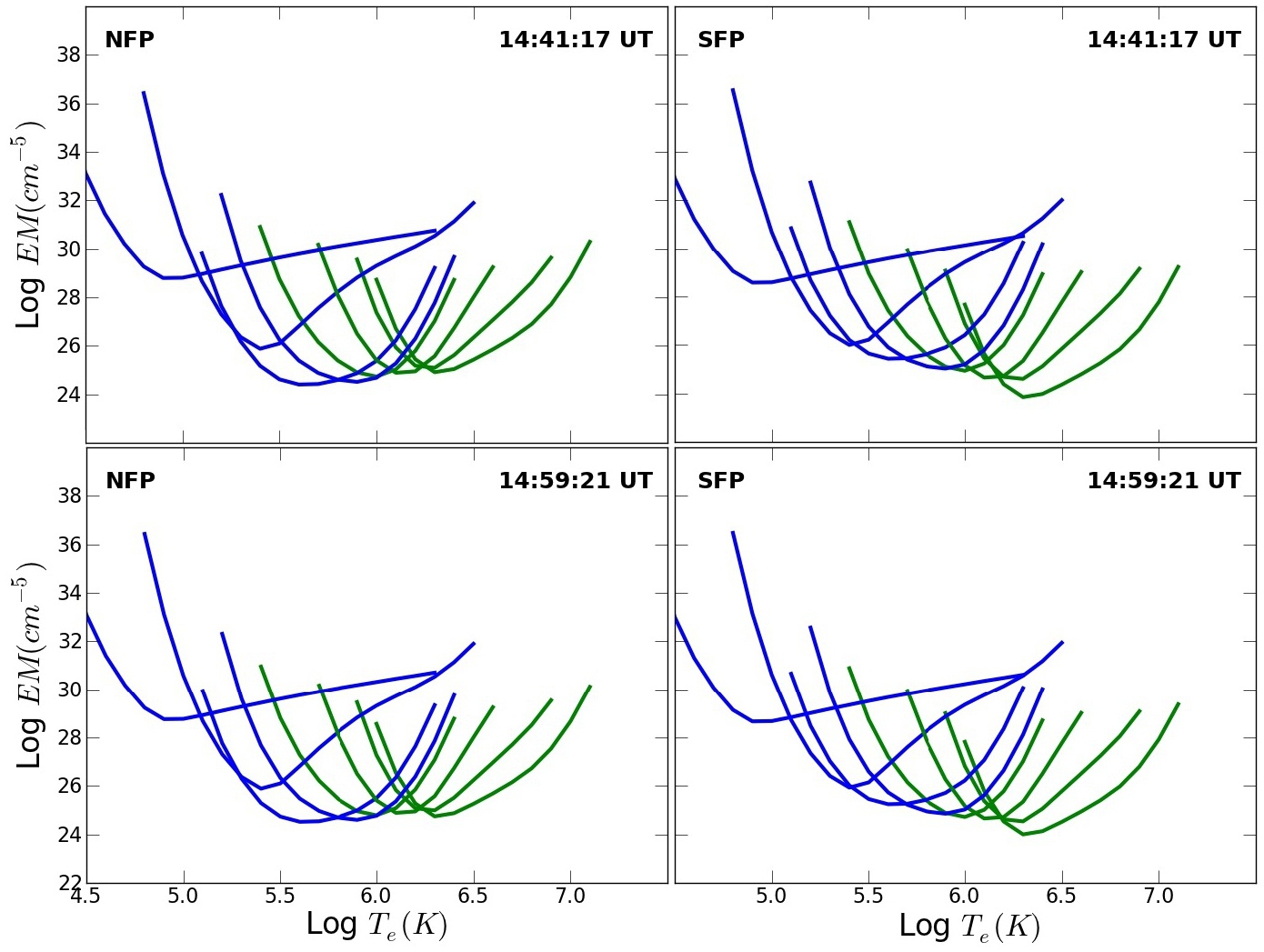}
   \caption{EM loci curves (blue and green represent emission lines with peak formation temperatures in the TR and corona, respectively) for observation times of 14:20:40 UT\,--\,14:59:21 UT (top to bottom, respectively) of the NFP and SFP regions, left and right columns, respectively, derived from EIS emission line intensities (Table~\ref{tbl:eis_elines}).
   }\label{fig:EMLS_1441_1459_footpoints}
\end{figure*}

In Figure~\ref{fig:EMLS_1441_1459_footpoints} we have provided EM loci curves for both footpoints as a function of observation time. It is observed, a distinctive isothermal component at $\log T$\,$\approx$\,6.2 ({\it i.e.}, NFP and SFP; Figure~\ref{fig:EMLS_1441_1459_footpoints}). We note, though not shown here, these results are indicative of the loop core's EM loci analysis as well. The isothermal component is expected given the hot loop's consistent visually bright nature both throughout our observational sequence and emission lines with such formation temperatures (Figure~\ref{fig:FFOV_1420_1459_snapshot}). Furthermore, the EM loci analysis indicates that the cool loop and its footpoints, peak formation temperatures $\le$\,1.0 MK, were non-isothermal during our analysis given its broad distribution of loci curves. These results, as well as the approximately constant TR EM distribution, are suggestive of unresolved structure \citep{Brooksetal2012ApJ} and are expected given the varying visual nature observed in emission lines formed at these temperatures (Figure~\ref{fig:FFOV_1420_1459_snapshot}).

The EM loci analysis also indicates the SFP is the site of condensation, based on observed differences between $EM$'s of the two footpoint regions ($\log T$'s\,$\le$\,6.0; Figure~\ref{fig:EMLS_1441_1459_footpoints}). This notion is expected when a single footpoint acts as the dominate energization site \citep{CraigMcClymont1986,McClymontCraig1987}. These results are further consistent with previous observations that the cool loop is filling from the SFP to the NFP. Moreover, they are consistent with both the evolution of flux and velocity observed in each of the loop regions as a function of temperature.


The evolution of the normalized magnetic flux density at both footpoint sites is shown in Figure~\ref{fig:densevol_NFP_SFP}. We note, the NFP and SFP sites corresponded to the positive and negative polarity magnetic flux, respectively. The positive magnetic flux density evolution, for intensities $>$\,20 G, is described in detail as follows ({\it i.e.}, NFP). The flux density was decreasing at a rate of $\approx$\,1\% min$^{-1}$ until the cool loop started to fill (14:41 UT). Magnetic flux density was increasing at a similar rate during 14:41 UT\,--\,14:59 UT, thus correlating with the complete filling of the cool loop. While the cool loop drained and returned to equilibrium the magnetic flux density was again decreasing at $\approx$\,1\% min$^{-1}$. For the SFP site, the magnetic flux density, intensities $<$\,$-$20 G, were approximately constant with minimal fluctuations, $\approx$\,$\pm$\,5\%, during the observation time frame studied (Figure~\ref{fig:densevol_NFP_SFP}). In Figure~\ref{fig:densevol_NFP_SFP} a magnetogram sequence centered on the NFP has been provided to show the major flux elements contributed to magnetic flux density measurements are contained and do not drift out of the FOV.

\section{Discussion}\label{sec:discussion}

We have presented observations of a cool loop ($\log T$\,$\le$\,6.0) directly below a thermally isolated hot coronal loop ($\log T$\,$\approx$\,6.2) recorded on 18 October 2011 by EIS near solar center ($\le$\,100$\arcsec$ in both the solar $x$ and $y$ directions). Our study, to best of our knowledge, provides the first observational evidence of plasma upflows in the presence of cool loops in quiet Sun regions. HMI LOS magnetic field observations were utilized to investigate the relationship between EUV flux evolutions and plasma motion compared to that of the underlying magnetic field. The intensity and velocity structure of the cool loop showed characteristics similar to those published by \citet{Tripathietal2012ApJ} for active region loops, and contrast those most typically found in the presence of such structures, {\it e.g.}, redshifts along and at the footpoints of these structures \citep[{\it e.g.},][]{DelZanna2008A&A,Tripathietal2009ApJ,Chesnyetal2012}. Both footpoints were predominately blueshifted throughout the upper TR and lower corona (5.8\,$\le$\,$\log T$\,$\le$\,6.2) during the cool loop's lifetime, except when filling began (14:41 UT). At this time a symmetrical flow was observed that corresponded with maximal upflow speeds in the NFP region ($v_\lambda$\,$\approx$\,60 km s$^{-1}$) peaking at $\log T$\,$\approx$\,5.9 and decreased with increasing temperature.
\begin{figure*}[!ht]
\centering
\includegraphics[scale=0.23]{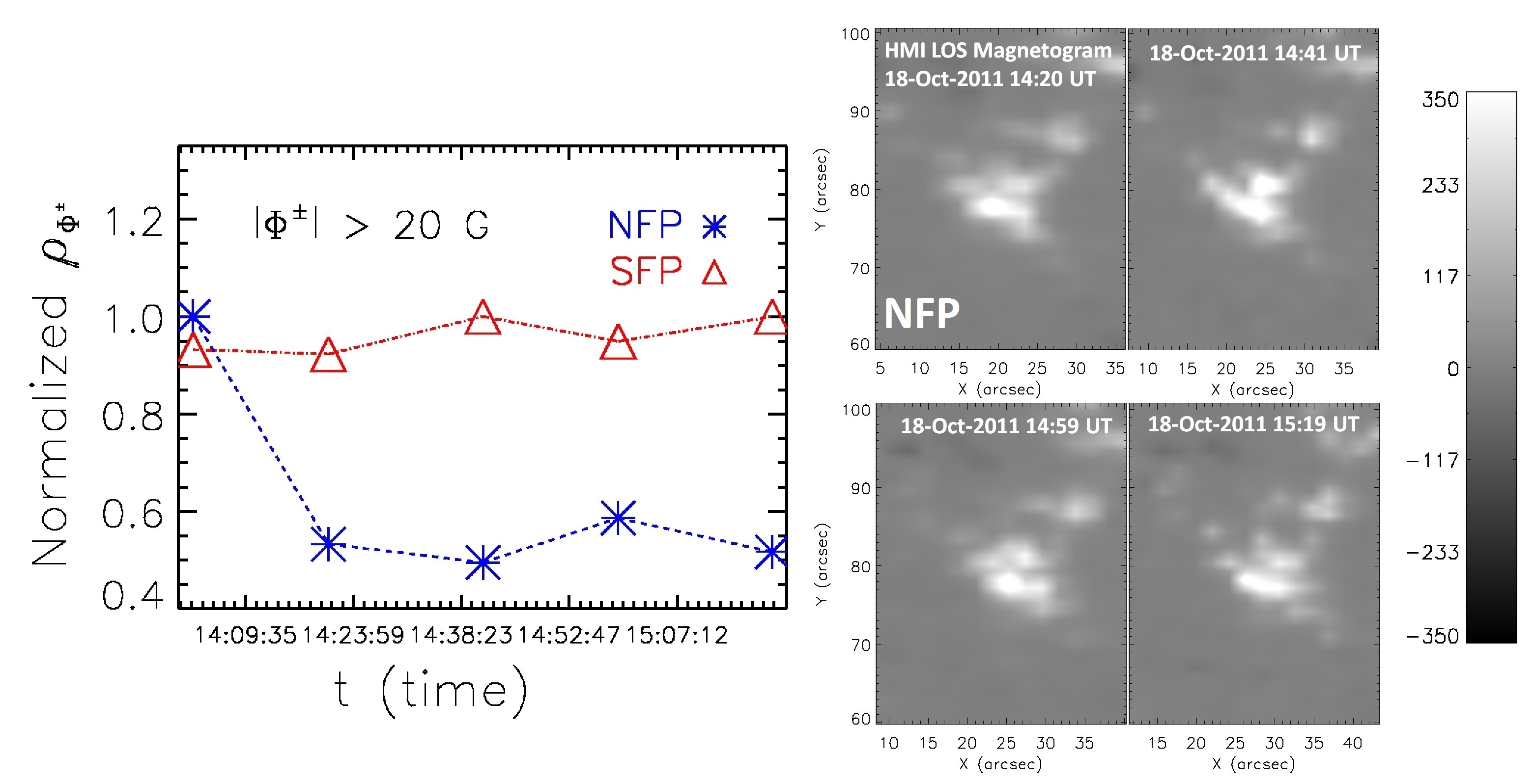}
   \caption{Right: evolution of the normalized magnetic surface flux density ($\rho_{\Phi^{\pm}}$; arbitrary units) for the NFP (asterisks, $\Phi^+$ $>$ 20 G) and SFP (triangles, $\Phi^-$ $<$ 20 G) regions over the observational time frame studied herein. Left: magnetogram temporal evolution of NFP region over the observational time frame of 14:20 UT\,--\,15:19 UT from top right to bottom right in clockwise fashion, respectively.}\label{fig:densevol_NFP_SFP}
\end{figure*}

As discussed in $\S$~\ref{sec:intro}, cool loops have often been considered to be a result of cooling and condensing coronal material which was heated impulsively over many strands at coronal heights. However, like the suggestions of \citet{Tripathietal2012ApJ}, our observations of plasma motions do not support models which predict upper TR and lower coronal emission lines that are dominated by redshifted emission. Our EM loci analysis indicates the presence of unresolved structure and non-isothermal plasma throughout the cooler layers of the atmosphere ($\log T$\,$\le$\,6.0; Figure~\ref{fig:EMLS_1441_1459_footpoints}). This result is expected in the presence of impulsive heating type events \citep{Brooksetal2012ApJ}. We explain these results by first noting that at the cool loop onset ({\it i.e.}, initial filling) non-steady symmetrical flows indicate an asymmetric loop structure \citep{Mariskaetal1982ApJ,CraigMcClymont1986,McClymontCraig1987}, while simultaneously plasma condensation is occurring at the SFP site (Figure~\ref{fig:EMLS_1441_1459_footpoints}). The runaway cooling observed $\approx$\,20 min later, 14:59 UT, in lower coronal and upper TR EUV images is then indicative of the movement of the condensation region to the less heated footpoint (i.e., NFP). \citet{CraigMcClymont1986} noted the temperature gradient of a condensing loop leg ({\it i.e.}, our SFP region) is shallower than that of the evaporating leg ({\it i.e.}, our NFP region), and as such is characterized by larger EMs. Therefore, using the minima of our footpoint EM distributions, over log T = 5.8 - 6.0 (Figure~\ref{fig:EMLS_1441_1459_footpoints}), a heating rate asymmetry of $\approx$\,2\% existed between the NFP and SFP regions. We point out, \citet{Mulleretal2003A&A} reported a 1\% energy asymmetry between loop legs dictates the draining direction which supports our observations of the condensation being driven from the SFP to the NFP. These results provide significant evidence that the catastrophic cooling event occurred from the loop's non-equilibrium state. Moreover, the non-equilibrium state formed when plasma condensation began in a single footpoint. Below we hypothesize on the mechanism responsible for initiating the condensation event.

\citet{Hegglandetal2009ApJ} suggest observations of solar atmospheric bi-directional jets are useful tools for probing the heights in which magnetic energy is being converted to thermal energy. As such, bi-directional jets provide unique diagnostic tools for constraining the heights of atmospheric heating. Inspecting our velocity versus temperature profiles prior to and at the onset of loop filling (Figure~\ref{fig:FPS_1420_LOSV_VS_TE}), a bi-directional jet is occurring between $\log T$\,$\approx$\,5.4\,--\,5.8 at the NFP site. Combining these observations with significant drops in magnetic surface flux density (Figure~\ref{fig:densevol_NFP_SFP}) occurring simultaneously both spatially with the NFP and temporally with peak plasma upflows, we find support that impulsive magnetic reconnection events between the photospheric footpoint and surrounding background field are the source of the jet. Consistencies of our observational reports to \citeauthor{Hegglandetal2009ApJ}'s (\citeyear{Hegglandetal2009ApJ}) TR simulated reconnection, lead us to suggest the reconnection event propelled a cool dense blob of plasma upwards along the field lines to the region of the SFP. Thereby, the SFP's sudden density enhancement (Figure~\ref{fig:1420_1459_densevol}), and most likely the conversion of magnetic wave energy to heat \citep{HollwegYang1988JGR,PoedtsGroof2004ESASP}, initiated plasma condensation. However, it cannot be ruled out that a dip in the magnetic field topology, at or very near the SFP, was responsible for the initiation of plasma condensation (Muller et al. 2003).

The previous discussion points to the fact that the cool loop was heated in a single footpoint ({\it i.e.}, SFP) which lead to a runaway cooling based on the non-equilibrium structure of the loop. The SFP heating event is considered to be low frequency in nature
\citep[{\it e.g.}, nanoflare;][]{Chittaetal2013arXiv}
as this explains our EM loci results (Figure~\ref{fig:EMLS_1441_1459_footpoints}) while remaining consistent with previous notions of cool loops \citep{Ugarteetal2009ApJ,Spadaroetal2003ApJ}. Our pervasive blueshifts immediately after catastrophic cooling, are indicative of plasma evaporation \citep{Tripathietal2012ApJ}, suggest the origin of the nanoflare storm was cooler regions of the solar atmosphere, particularly the upper TR. Moreover, these notions support footpoint heating scenarios, possibly at higher temperatures than typically considered, \citep{Tripathietal2012ApJ,Aschwandenetal2007ApJ}. Finally, our results also point to the fact that this whole process can be considered a domino effect as a direct result of TR magnetic reconnection in the opposing loop leg. Therefore, we have presented strong evidence supporting notions that coronal heating is not confined to only coronal temperatures, while tracing the origin of its magnetic energy conversion source.

It is worth noting, for $\log T$\,$\ge$\,6.2 light curves indicate footpoint heating of the hot loop in the NFP region, which peaked at $\approx$\,14:20 UT (Figure~\ref{fig:NFP_CORE_SFP_LCS}). Coronal EM loci provide evidence of condensation occurring simultaneously in time and spatial region (Figure~\ref{fig:EMLS_1441_1459_footpoints}). Using the techniques discussed previously, a heating asymmetry of $\approx$\,5\% was measured between the footpoints for temperatures  of $\log T$\,$\approx$\,6.2\,--\,6.4 (Figure~\ref{fig:EMLS_1441_1459_footpoints}). These results further support our suggestion of footpoint heating for the hot loop ({\it i.e.}, NFP region), as well as the reports of \citet{Tripathietal2012ApJ} and \citet{Aschwandenetal2007ApJ}. Finally, symmetrical flows in the hot loop at $\log T$\,$\approx$\,6.2 and 14:20 UT (Figure~\ref{fig:FPS_1420_LOSV_VS_TE}) provide evidence that the heating event was occurring at coronal heights with high enough frequency to maintain its visually and isothermally stable nature (Figures~\ref{fig:FFOV_1420_1459_snapshot} and \ref{fig:EMLS_1441_1459_footpoints}, respectively).

\section{Conclusion}\label{sec:conclusions}

This work has presented a play-by-play of the life cycle of a cool loop that emphasized the importance of the TR as the site of coronal heating. Our results have provided significant insight on the relationship between non-equilibrium structuring and condensation in cool plasma loops. It has also provided the first observational evidence, to best of our knowledge, of plasma upflows in the presence of cool loops in quiet Sun regions while supporting the findings of \citet{Tripathietal2012ApJ} for cool active region loops. We have built upon the work of \citet{Tripathietal2012ApJ} by including complimentary LOS magnetogram data and coronal density evolution. We conclude from the evolution of the underlying magnetic flux that observed TR upflows are an indication of magnetic reconnection at similar atmospheric heights.

We recognize, more observations are required to provide conclusive statements about the specifics of cool loop heating and the shared relationship between their continuous evolution and plasma condensation, which is planned for in a forthcoming paper. Moreover, these are required to better understand the cool loop's non-equilibrium nature as it relates to both asymmetrical and symmetrical flow patterns. Finally, plasma upflows in cool loops require further observational evidence to be used as constraints on the models proposed by \citet{Mulleretal2003A&A,Mulleretal2004A&A} to determine their validity. Lastly, we have not made suggestions on whether any heating connection exists between the cool and hot loop. However, significant coronal mass-influx peaking simultaneously with TR plasma upflows suggest such a connection exists. These notions give rise to questions of the mass and energy coupling relationship shared between these two loop structures, the TR and corona in general, and whether the high frequency heating events of the hot loop are a direct result of the conversion of magnetic to thermal energy at TR heights. \\~\\

The authors greatly appreciate the reviewers constructive comments on the manuscript. This research was supported by National Aeronautics and Space Administration (NASA) grant NNX-07AT01G and National Science Foundation (NSF) grant AST-0736479. Any opinions, findings and conclusions or recommendations expressed in this material are those of the author(s) and do not necessarily reflect the views of the NSF or NASA. The final publication is available at http://iopscience.iop.org/0004-637X/778/2/90/


\bibliographystyle{apj}
\bibliography{Fe8Loop_Col}

\end{document}